\documentclass[showkeys,showpacs,prd]{revtex4}
\usepackage{amsmath}
\usepackage{amssymb}
\usepackage{graphicx}
\usepackage{color}
\usepackage{ulem}
\usepackage{tikz}

\begin{document}

\title{
Three-point non-associative supersymmetry generalization
}

\author{
Vladimir Dzhunushaliev
}
\email{v.dzhunushaliev@gmail.com}
\affiliation{
Dept. Theor. and Nucl. Phys., KazNU, Almaty, 050040, Kazakhstan
}
\affiliation{
IETP, Al-Farabi KazNU, Almaty, 050040, Kazakhstan
}

\begin{abstract}
We consider a non-associative generalization of supersymmetry based on three-point associators like $\left[ Q_x, Q_y, Q_z \right]$ for $Q_{a, \dot a}$ supersymmetric generators. Such associators are connected with the products of $Q_{a, \dot a}$ and $x_{b \dot b}$. We: (a) calculate Jacobiators and show that the Jacobiators can be zero with some choice of corresponding coefficients in associators; (b) perform dimensional analysis for the coefficients in associators; (d) calculate some commutators involving coordinates and momentums; (e) estimate the weakness of non-associativity.
\end{abstract}


\maketitle

\section{Introduction}

Whether non-associativity could play a fundamental role in the formulation of physics is a question that has been raised from time to time. In Ref.~\cite{Mylonas:2012pg, Mylonas:2014aga, Mylonas:2013jha} the authors develop and investigate the quantization techniques for describing the nonassociative geometry probed by closed strings. In Ref.~\cite{Gunaydin:2013nqa} non-associative structures have appeared in the study of D-branes in curved backgrounds are investigated.

In Ref.\cite{Dzhunushaliev:2013by} we have discussed a possible generalization of supersymmetry with associator having four factors. Here we want to discuss a possible generalization of supersymmetry with associators having three factors (three-point associator). We will assume that such associators are connected with coordinates.

\section{Associative supersymmetric preliminaries}

In this section we would like to remember basic properties of the simplest supersymmetry algebra (see, for example, textbook \cite{Aitchison}). The most important for us is the anticommutator for $Q_a, Q_{\dot a}$ supersymmetry generators
\begin{equation}
  \left\{
     Q_a , Q_{\dot a}
  \right\} = Q_a Q_{\dot a} + Q_{\dot a} Q_a =
  2 \sigma^\mu_{a \dot a} P_\mu,
\label{2-10}
\end{equation}
and all other commutators and anticommutators are zero
\begin{eqnarray}
  \left\{
     Q_a , Q_b
  \right\} &=& \left\{
     Q_{\dot a} , Q_{\dot b}
  \right\} = 0,
\label{2-20} \\
  \left[ Q_a , P_\mu \right] &=& \left[ Q_{\dot a} , P_\mu \right] = 0,
\label{2-30}\\
  \left[ P_\mu , P_\nu \right] &=& 0,
\label{2-40}
\end{eqnarray}
here $P_\mu = -i \partial_\mu$ is the momentum operator;
$\mu = 0,1,2,3$; $a=1,2$;
$\dot a = \dot 1, \dot 2$. Pauli matrices
$\sigma^\mu_{a \dot a}, \sigma_\mu^{a \dot a}$ are defined in the standard way
\begin{eqnarray}
  \sigma^\mu_{a \dot a} &=& \left\{
    \left(
      \begin{array}{cc}
        1 & 0 \\
        0 & 1 \\
      \end{array}
    \right),
    \left(
    \begin{array}{cc}
        0 & 1 \\
        1 & 0 \\
      \end{array}
    \right),
    \left(
    \begin{array}{cc}
        0 & -i \\
        i & 0 \\
      \end{array}
    \right),
    \left(
    \begin{array}{cc}
        1 & 0 \\
        0 & -1 \\
      \end{array}
    \right)
  \right\}
\label{2-50}\\
  \sigma_\mu^{a \dot a} &=& \left\{
    \left(
      \begin{array}{cc}
        1 & 0 \\
        0 & 1 \\
      \end{array}
    \right),
    \left(
    \begin{array}{cc}
        0 & 1 \\
        1 & 0 \\
      \end{array}
    \right),
    \left(
    \begin{array}{cc}
        0 & i \\
        -i & 0 \\
      \end{array}
    \right),
    \left(
    \begin{array}{cc}
        1 & 0 \\
        0 & -1 \\
      \end{array}
    \right)
  \right\}
\label{2-60}
\end{eqnarray}
with orthogonality relations for Pauli matrixes
\begin{equation}\label{2-70}
  \sigma_\mu^{a \dot a} \sigma^\nu_{a \dot a} = 2 \delta_\mu^\nu, \quad
  \sigma_\mu^{a \dot a} \sigma^\mu_{b \dot b} =
  2 \delta_b^a \delta_{\dot b}^{\dot a}.
\end{equation}
Following the idea of Ref. \cite{Dzhunushaliev:2013by}, we want to show that one can generalize supersymmetry in such a way that the supergenerators $Q_a, Q_{\dot a}$ will become non-associative ones.

\section{Three-point non-associative generalization of the supersymmetry algebra}

In Ref. \cite{Dzhunushaliev:2013by} we have shown that it is possible to extend supersymmetry in such way that the generators
$Q_a, \bar Q_{\dot a}$ become non-associative ones. It was done by introducing a four-point associator in the following way
\begin{equation}\label{3-10}
  \left[ Q_a, Q_{\dot a}, \left( Q_b Q_{\dot b} \right) \right] =
  2 \zeta_0 \frac{\hbar}{\ell_0^2}
  \sigma^\mu_{a \dot a} \sigma^\nu_{b \dot b} M_{\mu \nu} +
  a_1 \epsilon_{ab} \epsilon_{\dot a \dot b} \mathbb I +
  \sigma^\mu_{a \dot a} \sigma^\nu_{b \dot b} a_2 P_\mu P_\nu +
  a_3 \eta_{\mu \nu} \sigma^\mu_{a \dot a} \sigma^\nu_{b \dot b} \mathbb I +
  a_4 \sigma^{\mu \nu}_{a b} \sigma^{\rho \tau}_{\dot a \dot b}
  M_{\mu \nu} M_{\rho \tau} + \cdots
  ,
\end{equation}
where $\zeta_0 = \pm i, \pm 1$, $\ell_0$ is some characteristic length; $a_{2,3,4}$ are complex numbers;
$
\sigma^{\mu \nu}_{a b} = \left[
  \sigma^\mu , \sigma^\nu
\right]_{ab}
$,
$
\sigma^{\mu \nu}_{\dot a \dot b} = \left[
  \sigma^\mu , \sigma^\nu
\right]_{\dot a \dot b}
$, and $\epsilon_{ab}, \epsilon_{\dot a \dot b}$ are the antisymmetric symbols; the coefficient $\hbar/\ell_0^2$ has been chosen so that the right- and left-hand sides of \eqref{3-10} have the same dimensions \cite{Dzhunushaliev:2015eva}.

Let us remember the definition of associator
\begin{equation}\label{3-20}
  \left[ A, B, C \right] = \left( A B \right) C - A \left( B C \right).
\end{equation}
Now we want to demonstrate that it is possible to introduce the  generalization of supersymmetry other than \eqref{3-10} based on a three-point associator. Let us define the following three-point associators:
\begin{eqnarray}
  \left[
    Q_{\dot a}, Q_a, Q_b
  \right] &=& \zeta_1 x_{a \dot a} Q_b +
  \zeta_2 x_{b \dot a} Q_a +
  \zeta_3 Q_a x_{b \dot a},
\label{3-48}\\
  \left[
    Q_a, Q_{\dot a}, Q_b
  \right] &=& \zeta_4 x_{a \dot a} Q_b +
  \zeta_5 Q_a x_{b \dot a} ,
\label{3-40}\\
  \left[
    Q_a, Q_b, Q_{\dot a}
  \right] &=& \zeta_6 Q_a x_{b \dot a} +
  \zeta_7 Q_b x_{a \dot a} +
  \zeta_8 x_{a \dot a} Q_b,
\label{3-30}\\
  \left[
    Q_a, Q_{\dot a}, Q_{\dot b}
  \right] &=& \xi_1 x_{a \dot a} Q_{\dot b} +
  \xi_2 x_{a \dot b} Q_{\dot a} +
  \xi_3 Q_{\dot a} x_{a \dot b} ,
\label{3-42}\\
  \left[
    Q_{\dot a}, Q_a, Q_{\dot b}
  \right] &=& \xi_4 x_{a \dot a} Q_{\dot b} +
  \xi_5 Q_{\dot a} x_{a \dot b} ,
\label{3-32}\\
  \left[
    Q_{\dot a}, Q_{\dot b}, Q_a
  \right] &=& \xi_6 Q_{\dot a} x_{a \dot b} +
  \xi_7 x_{a \dot a} Q_{\dot b} +
  \xi_8 Q_{\dot b} x_{a \dot a},
\label{3-45}\\
  \left[
    Q_a, Q_b, Q_c
  \right] &=& \left[
    Q_{\dot a}, Q_{\dot b}, Q_{\dot c}
  \right] = 0
\label{3-50}
\end{eqnarray}
where $a \neq b$; $\dot a \neq \dot b$; operators $x_{a \dot a}, x^\mu$ are connected with the relation
\begin{equation}\label{3-55}
  x_{a \dot a} = \sigma_{\mu a \dot a} x^\mu .
\end{equation}
We think that
\begin{equation}\label{3-56}
  \left| \zeta_i \right| = \begin{cases}
    \text{ either } & 0, \text{ for some } i \\
    \text{ or } & \zeta, \text{ for other values } i .
  \end{cases}
\end{equation}

\subsection{Dimensional analysis}

The simple dimensional analysis of equations \eqref{2-10}, \eqref{3-48} -- \eqref{3-50} shows that the dimensions of $\zeta_i$ and $\xi_i$ are
\begin{equation}\label{3-60}
  \left[ \zeta_i \right] = \left[ \xi_i \right] =
  \frac{\text{g}}{\text{s}} ,
  i=1, 2, \ldots , 8 .
\end{equation}
Again, as in Ref. \cite{Dzhunushaliev:2015eva}, we think that the relations \eqref{3-30} -- \eqref{3-50} should be quantum ones and consequently have to contain the Planck constant:
\begin{equation}\label{3-70}
  \zeta_i , \xi_i \propto \frac{\hbar}{\ell_0^2},
\end{equation}
where $\ell_0$ is some characteristic length, for example it can be done as $\ell_0^{-2} = \Lambda$ where $\Lambda$ is the cosmological constant.

\subsection{Jacobiators}

In this subsection we would like to calculate Jacobiator
\begin{equation}\label{4-1-10}
  J(x, y, z) = \left[
    \left[
      x, y
    \right] , z
  \right] +
  \left[
    \left[
      y, z
    \right] , x
  \right] +
  \left[
    \left[
      z, x
    \right] , y
  \right] = \left[ x, y, z \right] + \left[ y, z, x \right] +
  \left[ z, x, y \right] -
  \left[ x, z, y \right] - \left[ y, x, z \right] -
  \left[ z, y, x \right]
\end{equation}
where $x,y,z$ are $Q_{a, \dot a}$. Let us calculate Jacobiators
\begin{eqnarray}
  - J(Q_a, Q_{\dot a}, Q_b) &=& J(Q_a, Q_b, Q_{\dot a}) =
  J(Q_{\dot a}, Q_a, Q_b) =
\nonumber \\
  &&
  \left(
    \zeta_1 + \zeta_8 - \zeta_2 - \zeta_4
  \right) \left(
    x_{a \dot a} Q_b - x_{b \dot a} Q_a
  \right) + \left(
    \zeta_3 + \zeta_6 - \zeta_5 - \zeta_7
  \right) \left(
    Q_a x_{b \dot a} - Q_b x_{a \dot a}
  \right) ,
\label{4-1-20}\\
  - J(Q_{\dot a}, Q_a, Q_{\dot b}) &=& J(Q_a, Q_{\dot a}, Q_{\dot b}) =
  J(Q_{\dot a}, Q_{\dot b}, Q_a) =
\nonumber \\
  &&
  \left(
    \xi_6 + \xi_3 - \xi_5 - \xi_8
  \right) \left(
    Q_{\dot a} x_{a \dot b} - Q_{\dot b} x_{a \dot a}
  \right) + \left(
    \xi_4 + \xi_2 - \xi_1 - \xi_7
  \right) \left(
    x_{a \dot b} Q_{\dot a} - x_{a \dot a} Q_{\dot b}
  \right).
\label{4-1-30}
\end{eqnarray}
We see that the Jacobiators are zero either by
\begin{eqnarray}
  \zeta_1 + \zeta_8 - \zeta_2 - \zeta_4 &=& 0,
\label{4-1-40}\\
  \zeta_3 + \zeta_6 - \zeta_5 - \zeta_7 &=& 0,
\label{4-1-50}\\
  \xi_6 + \xi_3 - \xi_5 - \xi_8 &=& 0,
\label{4-1-60}\\
  \xi_4 + \xi_2 - \xi_1 - \xi_7 &=& 0
\label{4-1-70}
\end{eqnarray}
or by
\begin{eqnarray}
  x_{a \dot a} Q_b - x_{b \dot a} Q_a &=& 0,
\label{4-1-80}\\
  Q_a x_{b \dot a} - Q_b x_{a \dot a} &=& 0,
\label{4-1-90}\\
  Q_{\dot a} x_{a \dot b} - Q_{\dot b} x_{a \dot a} &=& 0,
\label{4-1-100}\\
  x_{a \dot b} Q_{\dot a} - x_{a \dot a} Q_{\dot b} &=& 0.
\label{4-1-110}
\end{eqnarray}
Probably the simplest case is
\begin{eqnarray}
  \zeta_2 &=& \zeta_3 = \zeta_7 = \zeta_8 = 0,
\label{4-1-120}\\
  \xi_2 &=& \xi_3 = \xi_7 = \xi_8 = 0,
\label{4-1-130}\\
  \left| \zeta_1 \right| &=& \left| \zeta_4 \right| =
  \left| \zeta_5 \right| = \left| \zeta_6 \right| =
  \left| \xi_1 \right| = \left| \xi_4 \right| =
  \left| \xi_5 \right| = \left| \xi_6 \right| =
  \frac{\hbar}{\ell_0^2} .
\label{4-1-140}
\end{eqnarray}
It means that
\begin{equation}\label{4-1-150}
  \zeta_i = \xi_i = \zeta_0 \frac{\hbar}{\ell_0^2}
\end{equation}
here $i = 1, 4, 5, 6$ and $\zeta_0 = \pm 1, \pm i$. In this case the associators \eqref{3-48} -- \eqref{3-50} are
\begin{eqnarray}
  \left[
    Q_{\dot a}, Q_a, Q_b
  \right] &=& \zeta_0 \frac{\hbar}{\ell_0^2} x_{a \dot a} Q_b ,
\label{4-1-160}\\
  \left[
    Q_a, Q_{\dot a}, Q_b
  \right] &=& \zeta_0 \frac{\hbar}{\ell_0^2} \left(
    x_{a \dot a} Q_b + Q_a x_{b \dot a} ,
  \right)
\label{4-1-170}\\
  \left[
    Q_a, Q_b, Q_{\dot a}
  \right] &=& \zeta_0 \frac{\hbar}{\ell_0^2} Q_a x_{b \dot a} ,
\label{4-1-180}\\
  \left[
    Q_a, Q_{\dot a}, Q_{\dot b}
  \right] &=& \zeta_0 \frac{\hbar}{\ell_0^2} x_{a \dot a} Q_{\dot b}  ,
\label{4-1-190}\\
  \left[
    Q_{\dot a}, Q_a, Q_{\dot b}
  \right] &=& \zeta_0 \frac{\hbar}{\ell_0^2} \left(
    x_{a \dot a} Q_{\dot b} + \xi_5 Q_{\dot a} x_{a \dot b}
  \right),
\label{4-1-200}\\
  \left[
    Q_{\dot a}, Q_{\dot b}, Q_a
  \right] &=& \zeta_0 \frac{\hbar}{\ell_0^2} Q_{\dot a} x_{a \dot b} ,
\label{4-1-210}\\
  \left[
    Q_a, Q_b, Q_c
  \right] &=& \left[
    Q_{\dot a}, Q_{\dot b}, Q_{\dot c}
  \right] = 0
\label{4-1-220}
\end{eqnarray}

\subsection{Commutator $\left[ x_\nu, p_\mu \right]$ in a non-associative form}
\label{commutativity}

Let us do the following transformations with 3-points associators \eqref{4-1-160}, \eqref{4-1-190}, \eqref{4-1-180} and \eqref{4-1-210}
\begin{equation}\label{4-2-10}
  \left[
    Q_{\dot a}, Q_a, Q_b
  \right] Q_{\dot b} +
  \left[
    Q_a, Q_{\dot a}, Q_{\dot b}
  \right] Q_b -
  Q_{\dot b} \left[
    Q_b, Q_a, Q_{\dot a}
  \right] -
  Q_b \left[
    Q_{\dot b}, Q_{\dot a}, Q_a
  \right] = \zeta_0 \frac{\hbar}{\ell_0^2}
  \sigma^\mu_{b \dot b} \sigma^\nu_{a \dot a}
  \left[ x_\nu, p_\mu \right] .
\end{equation}
We can use this expression for the estimation of weakness of non-associativity. Let us introduce dimensionless quantities
\begin{equation}\label{4-2-20}
  \tilde Q_{a, \dot a} = \frac{Q_{a, \dot a}}{Q_0},
  \tilde x^\mu = \frac{x^\mu}{l_{Pl}},
  \tilde p^\mu = \frac{p^\mu}{Q_0}
\end{equation}
where $Q_0 = \sqrt{\hbar c^3/G}$. Then \eqref{4-2-10} has the form
\begin{equation}\label{4-2-30}
  \left[
    \tilde Q_{\dot a}, \tilde Q_a, \tilde Q_b
  \right] \tilde Q_{\dot b} +
  \left[
    \tilde Q_a, \tilde Q_{\dot a}, \tilde Q_{\dot b}
  \right] \tilde Q_b -
  \tilde Q_{\dot b} \left[
    \tilde Q_b, \tilde Q_a, \tilde Q_{\dot a}
  \right] -
  \tilde Q_b \left[
    \tilde Q_{\dot b}, \tilde Q_{\dot a}, \tilde Q_a
  \right] = \zeta_0 \frac{l^2_{Pl}}{\ell_0^2}
  \sigma^\mu_{b \dot b} \sigma^\nu_{a \dot a}
  \left[ \tilde x_\nu, \tilde p_\mu \right]
\end{equation}
where $l_{Pl} = \sqrt{\hbar G/c^3}$ is the Planck length. If we us choose $\ell_0^{-2} = \Lambda$ ($\Lambda$ is the cosmological constant) then the coefficient $l^2_{Pl}/\ell_0^2$ on the RHS of \eqref{4-2-30} is $\approx 10^{-120}$. If $\left[ \tilde x_\nu, \tilde p_\mu \right] \sim 1$ then \eqref{4-2-30} shows that the non-associative expression on the LHS of \eqref{4-2-30} is extremely small $\approx 10^{-120}$.

The relation \eqref{4-2-10} can be inverted
\begin{equation}\label{4-2-40}
  \left[ x_\nu, p_\mu \right] = \frac{1}{4 \zeta_0} \frac{\ell_0^2}{\hbar}
  \sigma_\mu^{b \dot b} \sigma_\nu^{a \dot a} \left\{
  \left[
      Q_{\dot a}, Q_a, Q_b
    \right] Q_{\dot b} +
    \left[
      Q_a, Q_{\dot a}, Q_{\dot b}
    \right] Q_b -
    Q_{\dot b} \left[
      Q_b, Q_a, Q_{\dot a}
    \right] -
    Q_b \left[
      Q_{\dot b}, Q_{\dot a}, Q_a
    \right]
  \right\}
\end{equation}

\subsection{A weak version of coordinates non-commutativity}

In this subsection we calculate the anticommutator
\begin{equation}\label{4-10}
  \left\{
    \left[ Q_{\dot a}, Q_a, Q_c \right],
    \left[ Q_{\dot b}, Q_b, Q_d \right]
  \right\} =
  \zeta_0^2 \frac{\hbar^2}{\ell_0^4} \left[
    \left( x_{a \dot a} Q_c \right)
    \left( x_{b \dot b} Q_d \right) +
    \left( x_{b \dot b} Q_d \right)
    \left( x_{a \dot a} Q_c \right)
  \right].
\end{equation}
Let us assume that the operators $x_{a \dot a}$ are in an associative subalgebra of an non-associative algebra of $Q_{a, \dot a}$.
Then we can omit the brackets on the RHS of \eqref{4-10}. Introducing the commutator
\begin{equation}\label{4-20}
  \left[ Q_b, x_{a \dot a}, \right] = \alpha_{b, a \dot a},
\end{equation}
we  obtain
\begin{equation}\label{4-30}
  \frac{\ell_0^4}{\zeta_0^2 \hbar^2}
  \left\{
    \left[ Q_{\dot a}, Q_a, Q_c \right],
    \left[ Q_{\dot b}, Q_b, Q_d \right]
  \right\}  =
  x_{a \dot a} x_{b \dot b} Q_c Q_d +
  x_{b \dot b} x_{a \dot a} Q_d Q_c +
  x_{a \dot a} \alpha_{c, b \dot b} Q_d +
  x_{b \dot b} \alpha_{d, a \dot a} Q_c .
\end{equation}
Using the anticommutator \eqref{2-20}, we will have
\begin{equation}\label{4-40}
  \left[
    x_{a \dot a}, x_{b \dot b}
  \right] Q_c Q_d =
  \frac{\ell_0^4}{\zeta_0^2 \hbar^2}
  \left\{
    \left[ Q_{\dot a}, Q_a, Q_c \right],
    \left[ Q_{\dot b}, Q_b, Q_d \right]
  \right\} - x_{a \dot a} \alpha_{c, b \dot b} Q_d -
  x_{b \dot b} \alpha_{d, a \dot a} Q_c .
\end{equation}
Using the definition \eqref{3-55} and the inverse relations \eqref{2-70}, we will obtain a weak version of non-commutativity for coordinates
\begin{equation}\label{4-50}
  \left[
    x^\mu , x^\nu
  \right] Q_c Q_d = i \theta^{\mu \nu}_{c d},
\end{equation}
where the matrix $\theta^{\mu \nu}_{c d}$ is defined as
\begin{equation}\label{4-55}
  i \theta^{\mu \nu}_{c d} =
  \frac{1}{4} \sigma^{\mu a \dot a} \sigma^{\nu b \dot b} \left( \frac{\ell_0^4}{\zeta_0^2 \hbar^2}
    \left\{
      \left[ Q_y, Q_a, Q_{\dot a} \right],
      \left[ Q_z, Q_b, Q_{\dot b} \right]
    \right\} - x_{a \dot a} \alpha_{c, b \dot b} Q_d -
  x_{b \dot b} \alpha_{d, a \dot a} Q_c
  \right).
\end{equation}
The relation \eqref{4-50} have to be compared with the standard definition of non-commutativity of spacetime that can be encoded in the commutator of operators corresponding to spacetime coordinates \cite{Snyder:1946qz, Connes:2000ti}:
\begin{equation}\label{4-60}
  \left[
    x^\mu , x^\nu
  \right] = i \theta^{\mu \nu},
\end{equation}
where $\theta^{\mu \nu}$ is an antisymmetric matrix. We see that the relation \eqref{4-50} looks like a weak version of commutator for non-commutative coordinates in the consequence of the prefactor $Q_c Q_d$ in front of the commutator $\left[ x^\mu , x^\nu \right]$.

One can rewrite \eqref{4-10} in dimensionless form
\begin{equation}\label{4-70}
  \left\{
    \left[ \tilde Q_{\dot a}, \tilde Q_a, \tilde Q_c \right],
    \left[ \tilde Q_{\dot b}, \tilde Q_b, \tilde Q_d \right]
  \right\} =
  \zeta_0^2 \frac{l^4_{Pl}}{\ell_0^4} \left[
    \left( \tilde x_{a \dot a} \tilde Q_c \right)
    \left( \tilde x_{b \dot b} \tilde Q_d \right) +
    \left( \tilde x_{b \dot b} \tilde Q_d \right)
    \left( \tilde x_{a \dot a} \tilde Q_c \right)
  \right].
\end{equation}
Similarly to the previous subsection \ref{commutativity} we see that the RHS of \eqref{4-70} is extremely small $\approx 10^{-240}$.

\section{Discussion and conclusions}

Thus, here we have suggested a non-associative generalization of supersymmetry with three-point associators. We have shown that it is possible to choice the associators in such way that non-associative generators will satisfy to Jacobi identity. We have calculated the commutators
$\left [x_\mu, p_\nu \right]$ and $\left[ x^\mu, x^\nu \right]$. One question in this approach: is the operator $x^\mu$ from relations \eqref{3-30} and \eqref{3-40} the same or similar to quantized spacetime coordinates \`{a} la Snyder \cite{Snyder:1946qz} and non-commutative geometry \cite{Connes:2000ti}~? If yes, then such non-associative generalization of supersymmetry in some weak sense leads to an interesting connection with non-commutativity of coordinates. The dimensional analysis of parameter $\ell_0$ is done. The analysis gives rise to the conclusion that $\ell_0$ have to be connected with some characteristic length. For example, it can be the cosmological constant $\Lambda^{-1}$. In this case the manifestation of non-associativity is extremely small
$l_{Pl}^2/\ell_0^2 = l_{Pl}^2 \Lambda \approx 10^{-120}$.

\section*{Acknowledgements}

This work was supported by the Volkswagen Stiftung and by a grant No.~0263/PCF-14 in fundamental research in natural sciences by the Ministry of Education and Science of Kazakhstan. I am very grateful to V. Folomeev and A. Deriglazov for fruitful discussions and comments.

\end{document}